\documentclass[letterpaper,11pt]{article}
\pdfoutput=1 

\usepackage{jheppub} 
\usepackage[utf8]{inputenc}
\usepackage[T1]{fontenc} 
\usepackage[dvipsnames]{xcolor}
\usepackage{multirow}
\usepackage{slashed}
\usepackage[normalem]{ulem}
\usepackage{sidecap}
\usepackage{mathtools}
\usepackage{caption}
\usepackage{subcaption}
\usepackage[toc]{appendix}

\usepackage{amsmath}
\usepackage{graphicx}
\usepackage{fancyhdr}
\usepackage{multirow}
\usepackage{booktabs}
\usepackage{makecell}
\usepackage{tabulary}
\usepackage{feynmp}
\newcommand{\blue}[0]{\color{blue}}

\newcommand{\red}[0]{\color{red}}

\newcommand{\U}[1]{\mathrm{U}(1)_{\mathrm{#1}}}			

\title{Exploring the viability of pseudo Nambu-Goldstone boson as ultralight dark matter in a mass range relevant for strong gravity applications}

\author[a,b]{Ant\'onio~P.~Morais}
\author[c]{Vinícius Oliveira }
\author[d]{António Onofre}
\author[e]{Roman Pasechnik }
\author[f,g]{Rui Santos}

\affiliation[a]{Theoretical Physics Department, CERN, 1211 Geneva 23, Switzerland.}
\affiliation[b]{Departamento de F\'{i}sica da Universidade de Aveiro and Centre  for  Research  and  Development  in  Mathematics  and  Applications  (CIDMA),
	Campus de Santiago, 3810-183 Aveiro, Portugal.}
\affiliation[c]{Departamento de Física, Universidade Federal da Paraíba, Caixa Postal 5008, 58051-970, João Pessoa, PB, Brazil}
\affiliation[d]{Centro de Física das Universidades do Minho e do Porto (CF-UM-UP), Universidade do Minho, 4710-057 Braga, Portugal}
\affiliation[e]{Department of Physics, Lund University, SE-22100 Lund, Sweden}
\affiliation[f]{ISEL - Instituto Superior de Engenharia de Lisboa, Instituto Politécnico de Lisboa 1959-007 Lisboa, Portugal}
\affiliation[g]{Centro de Física Teórica e Computacional, Faculdade de Ciências,Universidade de Lisboa, Campo Grande, Edifício C8 1749-016 Lisboa, Portugal}

\emailAdd{aapmorais@ua.pt}
\emailAdd{vlbo@academico.ufpb.br}
\emailAdd{Antonio.Onofre@cern.ch}
\emailAdd{roman.pasechnik@hep.lu.se}
\emailAdd{rasantos@fc.ul.pt}

\usepackage{calligra}
\usepackage{calrsfs}
\DeclareMathAlphabet{\mathcalligra}{T1}{calligra}{m}{n}
\DeclareMathAlphabet{\pazocal}{OMS}{zplm}{m}{n}

\def\to{\rightarrow}

\def\figureautorefname~#1\null{Fig.\,#1\null}
\def\tableautorefname~#1\null{Tab.\,#1\null}

\def\equationautorefname~#1\null{Eq.\,(#1)\null}

\abstract{
We study a simple extension of the Standard Model featuring a dark sector with an ultralight pseudo Nambu-Goldstone boson as dark matter candidate. We focus on the mass range $\mathcal{O}(10^{-20} - 10^{-10})$ eV, relevant for strong gravity applications, and explore its production and evolution in the early Universe. The model is formulated in such a way that dark matter does not couple directly to photons or other Standard Model particles avoiding some of the most stringent cosmological bounds related to axion-like particles. In this work, two different scenarios are considered depending on whether dark matter is produced in a pre-inflationary or post-inflationary regime. We also discuss the effect from emergent topological defects such as cosmic strings and domain walls, and estimate the spectrum of stochastic gravitational waves produced by their decay, enabling to test the model at current and future gravitational-wave experiments.
}

\begin{document}

\begin{flushright}
CERN-TH-2023-070\\
\vskip1cm
\end{flushright}

\maketitle
\flushbottom

\section{Introduction}
\label{sec:intro}

There are remarkable evidences about the existence of a Dark Matter (DM) component in the Universe \cite{Bertone:2016nfn}, amounting to $27\%$ of its total energy density \cite{Planck:2018vyg,Drees:2018hzm}. A viable candidate must be electrically neutral (at least effectively \cite{Bogorad:2021uew}), cosmologically stable and non-relativistic during the period of matter-radiation equality. Among such candidates, Weakly Interacting Massive Particles (WIMPs) have been broadly studied and became widely popular due to their weak scale nature \cite{Bertone:2004pz,Steigman:2012nb,Arcadi:2017kky}. However, with the lack of both direct and indirect experimental evidence for WIMPs, the scientific community has been gradually increasing its attention to alternative candidates \cite{Arcadi:2017kky,Roszkowski:2017nbc}, such as Axion-Like-Particles, or ALPs for short \cite{Agrawal:2021dbo}. The latter can emerge in various contexts \cite{Kersten:2007vk,Peccei:1977hh,Peccei:1977ur,Dine:1981rt, Marsh:2015xka,Marsh:2017hbv} and provide an excellent solution for both thermal and non-thermal DM \cite{Battaglieri:2017aum,Arvanitaki:2014wva,Hui:2016ltb}.

Ultralight scalar particles such as ALPs are also well motivated in the context of Strong Gravity (SG). In general, relativistic ultralight bosons (scalars or vectors) can form self-gravitating lumps achieving a compactness comparable to that of black holes \cite{Kaup:1968zz,Ruffini:1969qy,Schunck:2003kk,Liebling:2012fv,Fox:2023aat} typically described in the context of general relativity minimally coupled to the bosonic field. A number of studies by the SG community developed in recent years, see \textit{e.g.}~\cite{Cardoso:2019rvt,Cardoso:2022nzc,CalderonBustillo:2020fyi,Sanchis-Gual:2022zsr,Chung-Jukko:2023cow,Acevedo:2022gug,Sanchis-Gual:2022mkk,Delgado:2020hwr,Sanchis-Gual:2021phr,Brito:2023fwr,Herdeiro:2023lze,Sengo:2022jif,CalderonBustillo:2022cja,Cardoso:2022vpj,Vicente:2022ivh,Annulli:2020lyc,Ikeda:2021pkb,Yuan:2021ebu,Ikeda:2018nhb,Cardoso:2018tly,Brito:2021war}, postulate that such particles account for, at least, a fraction of the observed DM abundance in the Universe. In this article, focusing on the case of a real pseudo Nambu-Goldstone Boson (pNGB), we explore under which circumstances such an assumption is valid. In particular, we discuss DM production mechanisms valid in the mass range of interest for SG scenarios, its relic abundance in the same region and the impact of cosmological and SG constraints. Last but not least, we investigate whether an ultralight pNBG can be associated to additional observables such as a Stochastic Gravitational Waves Background (SGWB) from the decay of long-lived topological defects \cite{Barman:2022yos,Dror:2019syi,Blasi:2020wpy,Fornal:2020esl,Samanta:2020cdk} and in which frequency range it can be tested.

A comprehensive study of simple Standard Model (SM) extensions relevant for SG applications was carried out in \cite{Freitas:2021cfi}. In this article we focus on what was denoted as Model 1 in Tab.~1 of \cite{Freitas:2021cfi}, where, in addition to the SM particle content, it also contains a new CP-even Higgs boson, $h_2$, and an ultralight real pNGB denoted as $\theta$. In the SG context the latter were studied \textit{e.g.}~in \cite{Ikeda:2021pkb,Yuan:2021ebu,Ikeda:2018nhb,Cardoso:2018tly,Brito:2021war}  and can lead to the formation of astrophysical objects such as oscillatons \cite{Seidel:1993zk,Bar:2018acw,Bar:2019bqz}. These are slightly time-dependent and decay, but can be very long lived, at least for the case of
spherical stars \cite{Page:2003rd}. With recent LIGO and Virgo data, searches for evidence of superradiance instabilities in the form of a SGWB \cite{Brito:2017wnc,Brito:2017zvb,Yuan:2022bem,Ng:2020ruv} have excluded the mass of real pNGB in the range $[1.3,17]\times 10^{-13}~\mathrm{eV}$. While this type of DM candidate has already been studied in other contexts \cite{Abe:2020ldj,Gross:2017dan,Masso:2004cv}, in this article we make the first study directly targeted at the mass range relevant for SG, \textit{i.e.}~$\mathcal{O}(10^{-20}- 10^{-10})$ eV. The presence of a $\mathbb{Z}_2$ discrete symmetry forbids direct couplings to photons of the form $\theta F^{\mu \nu} F_{\mu \nu}$ thus avoiding leading order cosmological bounds \cite{Cadamuro:2011fd} related to ALPs.

This article is organized as follows. In \autoref{sec:model} we revisit the key aspects of the model; in \autoref{sec: Dark_Matter} we discuss the production mechanisms viable in both pre and post-inflationary regimes; in \autoref{sec:pGW} we study the implications for the SGWB produced through the decay of topological defects; lastly, we draw our conclusions in \autoref{sec:conclusions}.

\section{The model}
\label{sec:model}

In this section, we revisit the basic details of Model 1 in \cite{Freitas:2021cfi} where it was presented in greater detail. This consists of an extension of the SM with an additional complex singlet $\phi$ charged under a global $\U{G}$ symmetry such that the scalar potential
\begin{equation}\label{eq:potential}
    V \left(H, \phi \right) = V_0 \left( H  \right) + \mu_\phi^2 \phi \phi^* + \frac{1}{2} \lambda_\phi |\phi \phi^*|^2 + \lambda_{H \phi} H^\dagger H \phi \phi^* + V_\text{soft}\,, 
\end{equation}
invariant under the transformation
\begin{equation}
    \phi \to e^{i \alpha} \phi \,.
\end{equation}
In \autoref{eq:potential} $V_0 (H)$ denotes the SM potential
\begin{equation}
    V_0(H) = \mu_H^2 H^\dagger H + \frac{1}{2} \lambda_H \left(H^\dagger H \right)^2\,,
\end{equation}
and $V_\text{soft}$ is a term that softly breaks the  $\U{G}$ global symmetry
\begin{equation}\label{eq:soft_breaking}
    V_\text{soft} =  \frac{1}{2} \mu_s^2 \left( \phi^2 + \phi^{* 2} \right)\,.
\end{equation}

The NGB field is described as a phase, $\theta$, of the field $\phi$,
\begin{equation}
    \phi = \frac{1}{\sqrt{2}} \left( \sigma +\nu_\sigma \right) e^{i\theta/\nu_\sigma}\,,
\end{equation}
where $\sigma$ represents radial quantum fluctuations about the vacuum expectation value (VEV) $\nu_\sigma$. The soft breaking term, \autoref{eq:soft_breaking}, is responsible for generating a mass to the NGB, proportional to $\mu_s$. As will be clear later, we are interested in the scenario where $\nu_\sigma \gg \nu_{h}$, where $\nu_{h} = 246~\mathrm{GeV}$ is the usual SM Higgs doublet VEV in the SM.
The mass matrix for the physical states $\left(h, \sigma, \theta \right)$ reads as
\begin{equation}
    \boldsymbol{M}^2 =
\left(
\begin{matrix}
v_h^2 \lambda_H & v_h v_\sigma \lambda_{H \phi} & 0\\
v_h v_\sigma & v_\sigma^2 \lambda_\phi & 0 \\
0 & 0 & -2 \mu_s^2
\end{matrix}
\right)\,,
\end{equation}
where $\boldsymbol{M}^2$ can be diagonalized by the orthogonal transformation
\begin{equation}
    \boldsymbol{m}^2 = \boldsymbol{O}^\dagger \boldsymbol{M}^2 \boldsymbol{O} = 
\left(
\begin{matrix}
m_{h_1}^2 & 0 & 0\\
0 & m_{h_2}^2 & 0 \\
0 & 0 &  m_\theta^2
\end{matrix}
\right)\,,
\end{equation}
with
\begin{equation} \label{eq:O}
    \boldsymbol{O} =
\left(
\begin{matrix}
\cos \alpha & \sin \alpha & 0\\
-\sin \alpha & \cos \alpha & 0 \\
0 & 0 & 1
\end{matrix}
\right)\,.
\end{equation}
The eigenvalues of the mass matrix are
\begin{equation}
    m_{h_{1,2}}^2 = \frac{1}{2} \left[ v_h^2 \lambda_H  + v_\sigma^2 \lambda_\phi \mp \sqrt{v_h^4 \lambda_H^2 + v_\sigma^4 \lambda_\phi^2 + 2 v_h^2 v^2_\sigma \left( 2 \lambda_{H \phi}^2 - \lambda_H \lambda_\phi \right)}      \right]\,,
\end{equation}
\begin{equation}
    m_\theta^2 = - 2 \mu_s^2 \, ,
\end{equation}
and the scalar mixing angle $\alpha$ satisfies
\begin{equation}
    \tan\left( 2 \alpha \right) = \left( \frac{2 \lambda_{H \phi} \, \nu_h \nu_{\sigma}}{\lambda_H \nu_h^2 - \lambda_\phi \nu_\sigma^2} \right)\,.
\end{equation}

In what follows, we fix $h_1$ to be the SM Higgs boson with a mass of $125~\mathrm{GeV}$. The couplings between $h_1$, $h_2$ and $\theta$ with the remaining SM particles are shown in Ref.~\cite{Freitas:2021cfi}. The mass of the second scalar ($h_2$) is a free parameter of the model. However, as discussed below, the hierarchy $\nu_\sigma \gg \nu_h$ is necessary in order to produce $\theta$ as DM, which results in $m_{h_2} \gg m_{h_1}$. This further results in an extremely suppressed coupling of $h_2$ and $\theta$ to the SM particles, preventing them from reaching thermal equilibrium in the early Universe. The physical quartic couplings read as
\begin{equation}
    \lambda_{\theta \theta \theta \theta} = - \frac{m_\theta^2}{6 \nu_\sigma^2}\,,
\end{equation}
where $\lambda_{\theta \theta \theta \theta} < 0$ indicates an attractive ultralight DM scenario, instead of a repulsive one, as discussed in \cite{Fan:2016rda}.

The current LHC constraints coming from precise Higgs couplings measurements impose an upper bound on the scalar mixing angle $\alpha$ of \autoref{eq:O}, which can be translated into $|\sin \alpha| \lesssim \mathcal{O}(0.1)$ \cite{ATLAS:2019qdc,Robens:2021rkl}.
Furthermore, due to the mass hierarchy $m_{h_2} \gg m_{h_1}$, the only contribution to the invisible decay width of the 125 GeV Higgs boson is the channel $h_1 \to \theta \theta$. In the limit where $m_\theta \sim 0~\mathrm{eV}$, the partial decay width of a Higgs boson for such a channel can be cast as \cite{Freitas:2021cfi}
\begin{equation}
    \Gamma_{h_1 \to \theta \theta} = \frac{1}{32 \pi^2} \frac{m_{h_1}^3}{\nu_\sigma^2} \sin ^2\alpha\,.
\end{equation}
Thus, the constraint on the scalar mixing angle discussed above imposes a lower bound on the VEV, of $\nu_\sigma > 75$ GeV, which is well bellow the ranges of $\nu_\sigma$ relevant for this work (see the discussion in Sec.~\ref{sec: Dark_Matter}), and the mixing between $h_1$ and $h_2$ can safely be ignored.

\section{Dark Matter}
\label{sec: Dark_Matter}

As discussed above, the mass range of interest in this article is $m_\theta \sim \mathcal{O} \left(10^{-20} -  10^{-10} \right)$ eV. However, it is known that DM masses as light as $10^{-24}$ eV are not excluded \cite{Diez-Tejedor:2017ivd}. Additionally, when $m_\theta \sim 10^{-33}$ eV, which is approximately the value of the Hubble constant today, $\theta$ behaves like dark energy \cite{Hlozek:2014lca}.

An ultralight scalar DM can be produced in the early Universe mainly through four different mechanisms \cite{Marsh:2015xka}: the misalignment mechanism \cite{Diez-Tejedor:2017ivd, Abbott:1982af, Preskill:1982cy,Dine:1982ah,Elahi:2022hgj,Bogorad:2021uew}, the decay of thermal relics \cite{Im:2019iwd}, thermally (via freeze-out) \cite{Carenza:2021ebx,Kolb:1990vq,Marsh:2015xka} and via the decay of topological defects \cite{Kawasaki:2014sqa}. The misalignment mechanism and the decay of topological defects are widely studied in the context of the QCD axion \cite{Preskill:1982cy, Abbott:1982af,Dine:1982ah, Marsh:2015xka,Hiramatsu:2012sc}, whose results are also applicable for a pNGB. If thermally produced, ultralight DM \cite{Dutra:2021lto,Patwardhan:2015kga} can become hot and jeopardize the period of structure formation \cite{Primack:2000iq}. It follows from $\nu_\sigma \gg \nu_\text{h}$ that the DM candidate considered in this article is neither produced by the decay of thermal relics nor via freeze-out as the couplings between the DM and SM particles are strongly suppressed. Other mechanisms, for instance freeze-in \cite{Hall:2009bx}, also provide negligible contributions due to tiny portal couplings between the Higgs boson and the dark sector.

In order to understand how $\theta$ is produced, let us briefly revisit the pNGB cosmology at the early Universe. Our DM candidate, $\theta$, emerges as a NGB when the global U$(1)_{\text{G}}$ symmetry is spontaneously broken, occurring at early times when $T \sim \nu_\sigma$ \cite{Kolb:1990vq,Marsh:2015xka,Hui:2021tkt}, and acquires a tiny mass via the soft term $V_\text{soft}$, becoming a pNGB. The spontaneous symmetry breaking (SSB) may happen before the end of inflation (we call it scenario I) or after it (scenario II), with very different consequences. We will now address in detail the two scenarios. 

The temperature of the Universe at the time of inflation is given by the Gibbons-Hawking expression \cite{Kolb:1990vq, Marsh:2015xka, PhysRevD.15.2738}
\begin{equation}\label{eq:TI}
    T_I = \frac{H_I}{2 \pi}\,,
\end{equation}
where $H_I$ is the inflationary Hubble parameter with an upper bound that comes from the Planck and BICEP2 \cite{BICEP2:2015nss, Marsh:2015xka} measurements,
\begin{equation}\label{eq:HI}
    H_I < 8.8 \times 10^{13} \ \text{GeV}\,.
\end{equation}
However, it is important to note that $H_I$ can be smaller and that its value depends on which inflationary landscape is being considered \cite{Kachru:2003aw, Martin:2013tda}.

When the global U$(1)_{\text{G}}$ symmetry is spontaneously broken $\theta$ can acquire any random initial value in the range $(0,2\pi]$. If SSB occurs before the end of inflation (scenario I) the initial values belong to different and causally disconnected patches of the Universe. Rapid expansion during inflation dilutes away the phase transition relics, and contributions from topological defects are washed out \cite{Diez-Tejedor:2017ivd,Marsh:2015xka,Bogorad:2021uew,Sikivie:2006ni}. In scenario I the observable Universe originates from a single causally connected region at the time of SSB featuring a single initial field value for the pNGB, whose relic abundance is dominated by the misalignment mechanism.

On the other hand, if SSB happens after the end of inflation (scenario II), the pNGB field will acquire different randomly chosen values in different causally disconnected regions of the Universe. In this case the formation of cosmic strings is expected \cite{Kolb:1990vq, Vilenkin:2000jqa}, that can later decay and produce DM. Additionally, if the potential has $N$ distinct degenerate minima in $\theta$, the formation of Domain Walls (DWs) \cite{Marsh:2015xka, Kolb:1990vq} will occur. When $N > 1$ stable DW are produced, as is our case with $N=2$. The energy density of stable DWs evolves slower than radiation and matter, and it can dominate the energy density of the Universe. This is the so-called DW problem \cite{Zeldovich:1974uw} which will be further discussed below. There is a large controversy about the appropriate approach to determine the production of ultralight particles (and, in particular, ALPs) from topological defects \cite{Arias:2012az}. In our analysis, we simply provide an estimate of such a contribution. For the case of scenario II, the relic abundance of $\theta$ is produced via the misalignment mechanism and through the decay of topological defects.

The soft-breaking term in the broken phase of $\phi$ can be cast as
\begin{equation}\label{eq:potentialZ2}
    V_{\text{soft}} = \frac{\mu_s^2}{2} \left(\sigma + \nu_\sigma \right)^2 \cos \left(2 \frac{\theta}{\nu_\sigma} \right)\,.
\end{equation}
It is clear that the potential has $N=2$ degenerate minima given by $\theta \to \theta + 2 \pi \nu_\sigma k/N$, where $k =0,1$, which makes the theory invariant under a remnant discrete $\mathbb{Z}_2$ symmetry. The latter emerges upon SSB of U$(1)_\text{G}$, with the pNGB transforming as $\theta \to -\theta$.
We now define the so-called misalignment angle \cite{Marsh:2015xka} as
\begin{equation}
    \Theta(x) \equiv \frac{\theta(x)}{\nu_\sigma \// 2}\,.
\end{equation}
We can expand the soft-breaking term (\autoref{eq:potentialZ2}) near to the minimum and for simplicity consider only the first term in the angle 
\begin{equation}
    V_{\rm soft} \simeq \frac{1}{8} m_\theta^2  \nu_\sigma^2  \Theta^2 \,.
\end{equation}
This means that as a first approximation we disregard the anharmonic terms, which become important only for large value of $\Theta$ \cite{Kawasaki:2014sqa,Visinelli:2009zm,PhysRevD.33.889,Bae:2008ue}, which is not our case as $|\Theta| \ll 1$ in order to represent the total DM density.

The equation of motion for $\Theta$ can be obtained by varying the action, $S= \int d^4x R^3 \mathcal{L}$, where $R$ is the scale factor, and for the Friedmann–Robertson–Walker (FRW) metric we obtain
\begin{equation}\label{eq:motionequation}
    \Ddot{\Theta} + 3H\Dot{\Theta} + m_\theta^2 \Theta = 0\,,
\end{equation}
which is an equation similar to the one describing the harmonic oscillator with time dependent friction. We note that $\theta$ is naturally stable, therefore the decay-type term, $\Gamma_\theta \dot{\Theta}$, is absent in \autoref{eq:motionequation} \cite{Kolb:1990vq}. Although $\sigma$ couples to $\theta$ in general, such an interaction is small enough to assume that the fields are approximately decoupled. This condition, together with the differential equation above, places us in the exact same scenario as the $\theta$ misalignment production in the axion scenario.

After the SSB of U$(1)_\text{G}$, the pNGB field assumes some initial non-zero value, $\theta_{\text{ini}}$. Then, as the initial value of $\theta$ is not necessarily aligned with the minimum of the potential, the pNGB field rolls down the potential and begins to oscillate coherently when $m_\theta \simeq 3 H \left(T_{\text{osc}}\right)$ \cite{Marsh:2015xka}, at some temperature given by\footnote{In this work, we assume the Standard Cosmology scenario, where the early Universe is dominated by radiation, with $H \simeq 1.66 \sqrt{g_*} T^2/M_{Pl}$, after inflation ($M_{Pl}$ is the Planck mass).}
\begin{equation}\label{eq: Tosc}
    T_\text{osc} \simeq 1.5 \text{ keV} \left( \frac{m_\theta}{10^{-20}\text{eV}}\right)^{1/2} \left( \frac{3.9
    }{g_*(T_\text{osc})}\right)^{1/4}\,,
\end{equation}
where $g_*(T_\text{osc})$ denotes the effective number of degrees of freedom at $T=T_\text{osc}$, with $T_\text{osc}$ greater than the temperature of the matter-radiation equality ($T_{\rm eq} \sim 1$ eV) \cite{Marsh:2015xka,Diez-Tejedor:2017ivd,Arias:2012az}. This in turn allows to extract a lower bound on $m_\theta$
\begin{equation}
    m_\theta \gtrsim 4.5 \times 10^{-27} \text{ eV}\,,
\end{equation}
which is consistent with the preferred range for SG applications.

The energy density originating from the coherent oscillation of $\theta$ at the minimum of the potential contributes to the final energy density of DM. This mechanism is called misalignment \cite{Marsh:2015xka} and occurs in both scenarios I and II. After this brief discussion about the pNGB cosmology, we will now estimate the final relic density of $\theta$.

\subsection{Pre-inflationary scenario}

The first scenario to discuss considers SSB of the U$(1)_\text{G}$ symmetry before the end of inflation. As we mentioned above, in this case $\theta$ will be predominantly produced via the misalignment mechanism and isocurvature perturbations will appear \cite{Marsh:2015xka}. The Cosmic Microwave Background (CMB) constrains the amplitude of the isocurvature perturbations, imposing a limit on the inflationary Hubble parameter $H_I$ as a function of the VEV
\begin{equation}
    \frac{H_I}{\nu_\sigma} \lesssim 3 \times 10^{-5}\,,
\end{equation}
implying that $\nu_\sigma \gtrsim 3.3 \times 10^{16}$ GeV, for $H_I = 10^{12}$ GeV. Nonetheless, as the Hubble scale at inflation is unknown a certain degree of freedom is still allowed.

In order to determine the relic abundance of the $\theta$ field we need to understand how its energy density has evolved during the early stages of the Universe evolution. The energy density of $\theta$ coherent oscillations can be obtained from the energy momentum tensor \cite{Marsh:2015xka} resulting in
\begin{equation}\label{eq:Energy_density}
    \rho_\theta^\text{mis} = \left( \frac{\nu_\sigma}{2} \right)^2 \left[\frac{\Dot{\Theta}^2}{2}  + \frac{1}{2} m_\theta^2 \Theta^2 \right]\,.
\end{equation}
Notice that \autoref{eq:motionequation} contains two real roots, implying that the system behaves as an overdamped oscillator. When the global U$(1)_\text{G}$ symmetry is spontaneously broken, $\Theta$ acquires an initial random value, $\Theta_i$. At time scales before $T_{\text{osc}}$, that is when $T \gg T_{\text{osc}}$, we have $\dot{\Theta} \simeq 0$, and therefore $\Theta$ is approximately constant. At a later time, when $T \ll T_{\text{osc}}$, the field rolls down the potential and begins to oscillate. Since $\Theta$ is approximately constant until $T_{\text{osc}}$, we can conclude that $\rho_\theta^\text{mis}$, in \autoref{eq:Energy_density}, is also approximately constant during this period \cite{Kolb:1990vq,Marsh:2015xka}. Therefore, we can take the energy density at $T_\text{osc}$ to be
\begin{equation}
    \rho_\theta^\text{mis} (R_\text{osc}) \simeq \frac{m_\theta^2 \Theta_i^2}{2} \left( \frac{\nu_\sigma}{2}\right)^2\,,
\end{equation}
where $\Theta_i$ describes the initial misalignment angle when the field starts to oscillate. After $T_\text{osc}$ the angle decreases behaving as non-relativistic matter, $\rho_\theta^\text{mis} \propto R^{-3}$.
The energy density of $\theta$ at some instant after $T_\text{osc}$ can be calculated using the redshift of the energy density in that period \cite{Marsh:2015xka,Elahi:2022hgj}
\begin{equation}
    \rho^\text{mis}_\theta(R) = \rho_\theta^\text{mis} \left( R_\text{osc} \right) \left(\frac{R_\text{osc}}{R} \right)^3 \simeq   \frac{m_\theta^2 \Theta_i^2}{2} \left( \frac{\nu_\sigma}{2}\right)^2 \left(\frac{R_\text{osc}}{R} \right)^3\,.
\end{equation}
As alluded above, notice that for a given mass, $m_\theta$, the energy density of $\theta$ is controlled by its initial value, vanishing in the case of $\Theta_i = 0$.
Using the definition of entropy density and assuming that it is constant, one obtains
\begin{equation}\label{eq: scale_factor}
    \left(\frac{R_{\rm osc}}{R_0} \right)^3 = \frac{g_s(T_0)}{g_s(T_{\rm osc})} \left( \frac{T_0}{T_{\rm osc}}\right)^3\,,
\end{equation}
where $g_s(T_0) = 3.91$ and $T_0 = 2.3 \times 10^{-4}$ eV is the temperature today \cite{ParticleDataGroup:2020ssz}.

The DM relic density is defined as
\begin{equation}
    \Omega_\theta h^2 = \frac{\rho_\theta^\text{mis} (R_0)}{\rho_{\rm crit}/h^2}\,.
\end{equation}
Taking $\rho_{\rm crit} = 1.05 \times 10^{-5} \ h^2 \ \text{GeV}/\text{cm}^3 \simeq 8.15 \times 10^{-47} \ h^2 \ $ GeV$^4$ \cite{ParticleDataGroup:2020ssz}, one can write
\begin{equation}\label{eq: AbundanceI}
    \Omega_\theta h^2 = 0.11 \left( \frac{m_\theta}{10^{-20} \text{ eV}}\right)^{1/2} \left( \frac{\nu_\sigma}{10^{17} \text{ GeV}} \right)^2  \left( \frac{\Theta_i}{1.5 \times 10^{-1}}\right)^2 \mathcal{F}\left(T_\text{osc} \right) \,,
\end{equation}
where $\mathcal{F}\left(T_\text{osc} \right) \equiv \left(\frac{3.91}{g_s(T_{\rm osc})} \right) \left( \frac{g_*(T_{\rm osc})}{3.4} \right)^{3/4}$ varies in the range from $1$ to $\sim 0.3$. This results in an upper limit on $\Theta_i$ compatible with $\Omega_\theta h^2 \leq 0.11$ that reads as
\begin{equation}
    \Theta_i \leq 0.15 \left(\frac{10^{-20} \text{eV}}{m_\theta} \right)^{1/4} \left(\frac{10^{17}\text{GeV}}{\nu_\sigma} \right) \mathcal{F}\left(T_\text{osc} \right) ^{-1/2}\,.
\end{equation}

\begin{figure}
    \centering
    \includegraphics[width=\columnwidth]{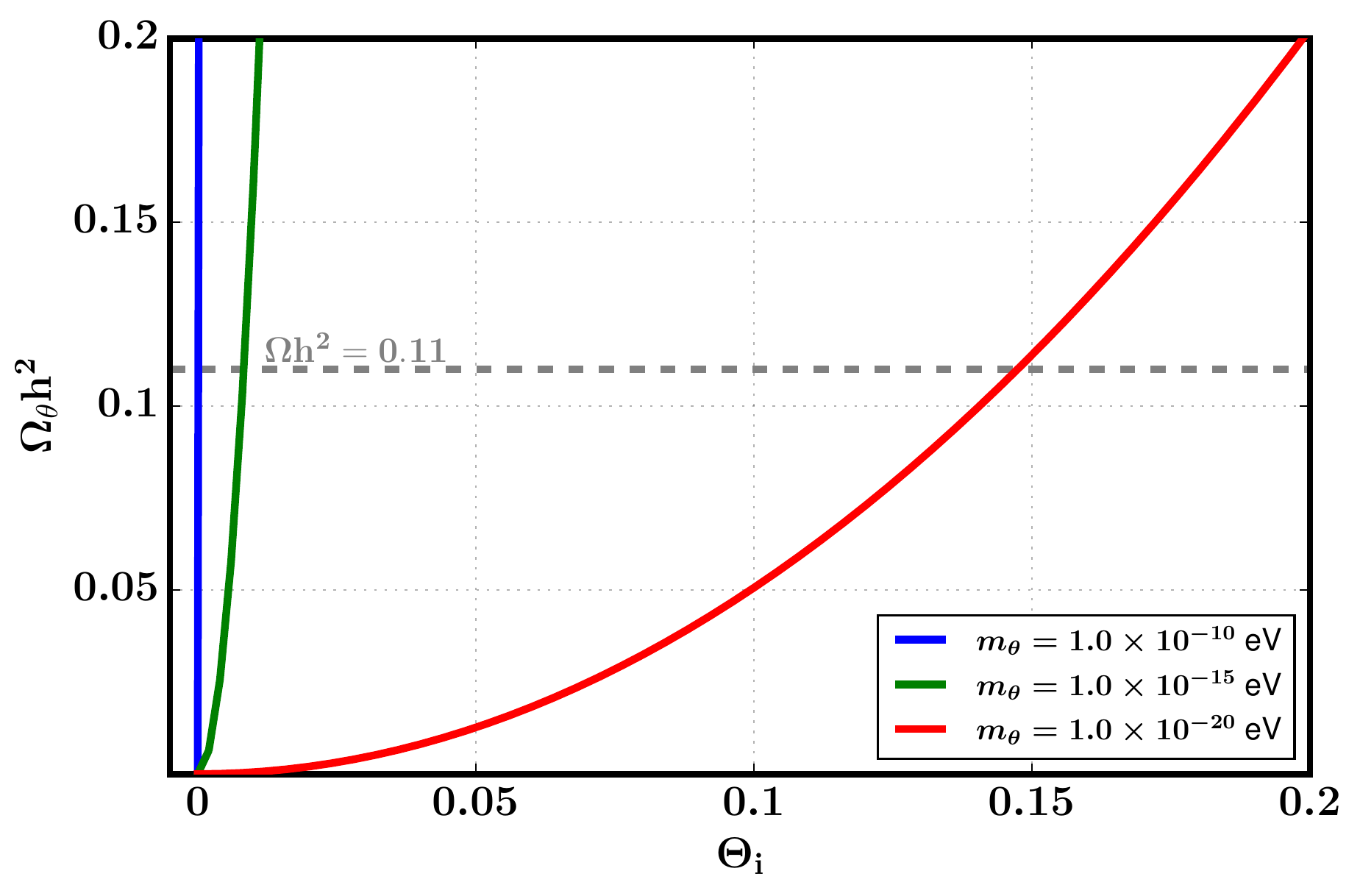}
    \caption{Abundance of ultralight DM for the case where the SSB occurs before the end of inflation, for {\blue $m_\theta= 10^{-10}$ eV}, {\color{ForestGreen} $10^{-15}$ eV} and {\red $10^{-20}$ eV} for blue, green and red curves, respectively. Here, we fixed $\nu_\sigma = 10^{17}$ GeV.}
    \label{fig:abudanceI}
\end{figure}

In \autoref{fig:abudanceI} we show the solution of \autoref{eq: AbundanceI} for three representative values of the DM mass, for instance, {\blue $m_\theta= 10^{-10}$ eV (blue curve)}, {\color{ForestGreen} $10^{-15}$ eV (green curve)} and {\red $10^{-20}$ eV (red curve)}. As can be observed, for the mass range of interest in our discussion, it is always possible to saturate the DM relic abundance with the $\theta$ field by fixing the initial value of the misalignment angle $\Theta_i = 4.7 \times 10^{-4}, 8.4 \times 10^{-3}$ and $1.5 \times 10^{-1}$ for  $m_\theta= 10^{-10}$ eV, $10^{-15}$ eV and $10^{-20}$ eV, respectively.

\subsection{Post-inflationary scenario}

Let us now consider the scenario where the global symmetry breaking occurs after the end of inflation, \textit{i.e.}~$T_I > \nu_\sigma$. As previously discussed, in this case DM is produced through the decay of topological defects (cosmic strings and DWs) \cite{Vilenkin:1984ib} besides the misalignment mechanism that is also present. Cosmic strings emerge when the U$(1)_\text{G}$ gauge symmetry is spontaneously broken at $T \sim \nu_\sigma$. After that, a potential with degenerate minima and a $\mathbb{Z}_2$ discrete symmetry arises. This symmetry, when spontaneously broken at $T_\text{osc}$, produces DWs.

The approach to obtain the contribution from misalignment is in many aspects identical to what is described above for scenario I. However, a few differences regarding the value of $\Theta_i$ need to be accounted for. In particular, when the U$(1)_\text{G}$ symmetry is spontaneously broken, the Universe is divided into several causally disconnected patches, each of which has an independent value of $\Theta_i$. Therefore, $\Theta_i$ is no longer a free parameter as before, and it is reasonable to consider the average of an uniform distribution \cite{Marsh:2015xka, Kolb:1990vq, Reig:2019sok}, $\Theta_i^2 \to \langle \Theta_i^2 \rangle$, where, neglecting the anharmonic effect, one has
\begin{equation}\label{eq:average_theta}
    \langle \Theta_i^2 \rangle = \frac{1}{2\pi}\int_{-\pi}^{\pi} d\Theta_i \Theta_i^2 = \frac{\pi^2}{3}\,.
\end{equation}

Notice that from $T_I > \nu_\sigma$ and using \autoref{eq:TI} and \autoref{eq:HI} one can obtain the upper limit on the $\U{G}$ breaking scale as $\nu_\sigma \lesssim 1.4 \times 10^{13}$ GeV.
Thus, the contribution from the misalignment mechanism for this scenario is obtained from \autoref{eq: AbundanceI} and \autoref{eq:average_theta} and can be written as
\begin{equation}\label{eq: AbundanceII}
    \Omega_\theta^{\rm mis} h^2 \simeq 1.6 \times 10^{-15} \left( \frac{m_\theta}{10^{-20} \text{ eV}}\right)^{1/2} \left( \frac{\nu_\sigma}{10^{9} \text{ GeV}} \right)^2  \mathcal{F}\left(T_\text{osc} \right)\,.
\end{equation}
With the considered values for the ultralight scalar mass in a range relevant for compact astrophysical objects, the misalignment contribution to the total relic density of $\theta$ is extremely suppressed as previously noticed in other works \cite{Reig:2019sok, Kolb:1990vq}. Therefore, one must closely inspect the contribution from other sources such as the decay products of topological defects.

The global U$(1)_\text{G}$ symmetry breaks spontaneously when $T \sim \nu_\sigma$ which induces the formation of cosmic strings \cite{Kawasaki:2014sqa,Davis:1986xc}. Such strings decay into pNGBs \cite{Davis:1985pt,Hiramatsu:2010yu,Kolb:1990vq} at $m_\theta t_{\text{str}} \sim 1$ and the energy stored in the cosmic strings is released dominantly into DM. The evolution of the energy density of these strings is described by the scaling solution
\begin{equation}
    \rho^\text{string}(t) = \xi \frac{\mu_\text{string}}{t^2}\,,
\end{equation}
where $\xi$ represents the number of strings in a Hubble volume. We will assume $\xi \sim 1$ \cite{Hiramatsu:2012gg}, and
\begin{equation}
    \mu_{\text{string}} = \pi \nu_\sigma^2 \log\left(\frac{t}{d} \right) 
\end{equation}
is the energy of a string per unit length, with $t = 1/H$ and $d \sim 1/m_{h_2}$ the width of the string \cite{Hiramatsu:2010yu}.

When $H \sim m_\theta$ the $\mathbb{Z}_2$ discrete symmetry, which emerges after SSB of U$(1)_\text{G}$, breaks and leads to the formation of stable DWs \cite{Reig:2019sok, Vilenkin:2000jqa}. In order to avoid the DW problem \cite{Zeldovich:1974uw}, we consider an additional generic term in the potential, $\delta V$, which slightly breaks the discrete $\mathbb{Z}_2$ symmetry and lifts the degeneracy between the two $\mathbb{Z}_2$ symmetric vacua evident in \autoref{eq:potentialZ2}. Additionally, we require that $V_\text{soft} > \delta V$ since the bias term can contribute to the $\theta$ mass.  The new $\delta V$ term produces a pressure on the walls and eventually annihilates them. The total potential (\autoref{eq:potential}) can then the rewritten
\begin{equation}\label{eq:potential_Z2_breaks}
    V \left(H, \phi \right) = V_0 \left( H  \right) + \mu_\phi^2 \phi \phi^* + \frac{1}{2} \lambda_\phi |\phi \phi^*|^2 + \lambda_{H \phi} H^\dagger H \phi \phi^* + V_{\rm soft} + \delta V\,.
\end{equation}
The $\delta V$ contribution was first proposed by Sikivie in \cite{PhysRevLett.48.1156} and represents a pressure term \cite{Hiramatsu:2010yn,Kawasaki:2014sqa,PhysRevD.39.1558,Caputo:2019wsd}. Here we are not concerned about an exact expression for $\delta V$, as our focus is rather on the difference in the values of the potential calculated at each minimum $\Delta V \simeq \delta V$ \cite{Gorghetto:2022ikz,Hiramatsu:2012sc}. A detailed discussion about DWs dynamics with pressure terms can be found in \cite{Larsson:1996sp}.

When the DWs are produced each cosmic string becomes attached to the DWs and a string-wall network emerges. As for the case of strings, numerical studies show that the energy density of the $\mathbb{Z}_2$ DWs is described by the scaling solution \cite{Caputo:2019wsd}
\begin{equation}\label{eq:DW_ED}
    \rho^\text{wall}(t) = A \frac{ \sigma_{\text{wall}}}{t}\,.
\end{equation}
where $A \simeq 0.8 \pm 0.1$ is the area parameter \cite{Hiramatsu:2013qaa} and $\sigma_{\text{wall}}$ is the tension of the DW \cite{Hiramatsu:2012sc,Saikawa:2017hiv}
\begin{equation}\label{eq: tension_wall}
    \sigma_{\text{wall}} = 2 m_\theta \nu_\sigma^2\,.
\end{equation}
As mentioned above, the pressure term lifts the degeneracy between the two vacua, producing a volume pressure $p_V \sim \Delta V$ on the DWs, proportional to the difference between the minima. Besides the volume pressure, DWs are also affected by a tension force $p_T \sim  \sigma_{\text{wall}} / t$ such that their decay occurs when $p_V \sim p_T$, which implies  \cite{Reig:2019sok}
\begin{equation}\label{eq: Hubble_decay}
    H_{\text{decay}} =  \frac{\Delta V}{A \sigma_\text{wall}}\,.
\end{equation}
As DWs predominantly decay into DM producing the gravitational waves (GWs), there is no conflict with the Big-Bang Nucleosynthesis (BBN) predictions. Hence, we only have to require that the DWs decay before the matter-radiation equality $H_\text{decay} > H_\text{eq}$ ($T_\text{eq} \sim 1 $ eV), which yields the condition  \cite{Caputo:2019wsd,Reig:2019sok,Marsh:2015xka}
\begin{equation}
     \Delta V > 2.5 \times 10^{-35}\text{ MeV}^4\, A  \left( \frac{m_\theta}{10^{-20} \text{ eV}} \right) \left( \frac{\nu_\sigma}{10^{9} \text{ GeV}} \right)^2 \,.
\end{equation}
The temperature at which the DWs decay is given by
\begin{eqnarray}
    T_{\text{decay}}  &\simeq & 1.98 \times 10^{8} \text{ GeV } A^{-1/2} \left( \frac{g_*(T_{\text{decay}})}{10}\right)^{-1/4} \left(\frac{m_\theta}{10^{-20} \text{eV}} \right)^{-1/2} \nonumber \\  &&
    \left( \frac{v_\sigma}{10^9 \text{GeV}} \right)^{-1/2} \left( \frac{\Delta V}{\text{MeV}^4} \right)^{1/2}\,.
\end{eqnarray}
Another lower bound on $\Delta V$ arises from requiring that the DWs should decay before they dominate the energy density of the Universe, which happens when  $t_r \sim (G \sigma_\text{wall})^{-1}$, where $G = 6.7 \times 10^{-39}$ GeV$^{-2}$ is the gravitational constant. Therefore, we require
\begin{equation}
    \frac{1}{G A \sigma_\text{wall}} \gg H^{-1}_\text{decay}\,,
\end{equation}
and using \autoref{eq: tension_wall} and \autoref{eq: Hubble_decay} we obtain 
\begin{equation}
    \Delta V \gg 2.68  \times 10^{-48} \text{ MeV}^4 \, A^2
    \left(\frac{m_\theta}{10^{-20}\text{ eV}} \right)^2  \left(\frac{\nu_\sigma}{10^{9}\text{ GeV}} \right)^4\,.
\end{equation}

To estimate the energy density of DM produced through the decay of the DWs we can redshift \autoref{eq:DW_ED}, where the produced DM behaves as matter (non-relativistic),
\begin{equation}
    \rho_\theta^{\text{wall}}(t_0) = \rho^\text{wall}(t_\text{decay}) \left( \frac{R(t_\text{decay})}{R(t_0)}\right)^3\, ,
\end{equation}
where
\begin{equation}
    \rho^\text{wall}(t_\text{decay}) = 2 \sigma_\text{wall} H_\text{decay}\, .
\end{equation}
Putting everything together, the relic density of pNGB produced via the decay of DWs is given by
\begin{equation}
    \Omega_\theta^\text{wall} h^2 \simeq 1.5 \times 10^{-29} A^{1/2} \left( \frac{g_*(T_\text{decay})}{10} \right)^{-1} \left( \frac{m_\theta}{10^{-20} \text{ eV}} \right)^{3/2} \left( \frac{\nu_\sigma}{10^9 \text{ GeV}} \right)^{3} \left( \frac{\Delta V}{\text{MeV}^{4}} \right)^{-1/2}\,.
\end{equation}
Last but not least, the total relic density of the ultralight scalar in a regime where the global $\U{G}$ is broken in a post-inflationary era reads as
\begin{equation}
    \Omega_\theta h^2 \simeq \Omega_\theta^\text{mis}h^2 + \Omega_\theta^\text{wall}h^2 \,,
\end{equation}
where the cosmic strings contribution is negligible, as shown in \cite{Kawasaki:2014sqa}, and thus neglected in the remainder of this analysis.
\begin{figure}
    \centering
    \includegraphics[width=\columnwidth]{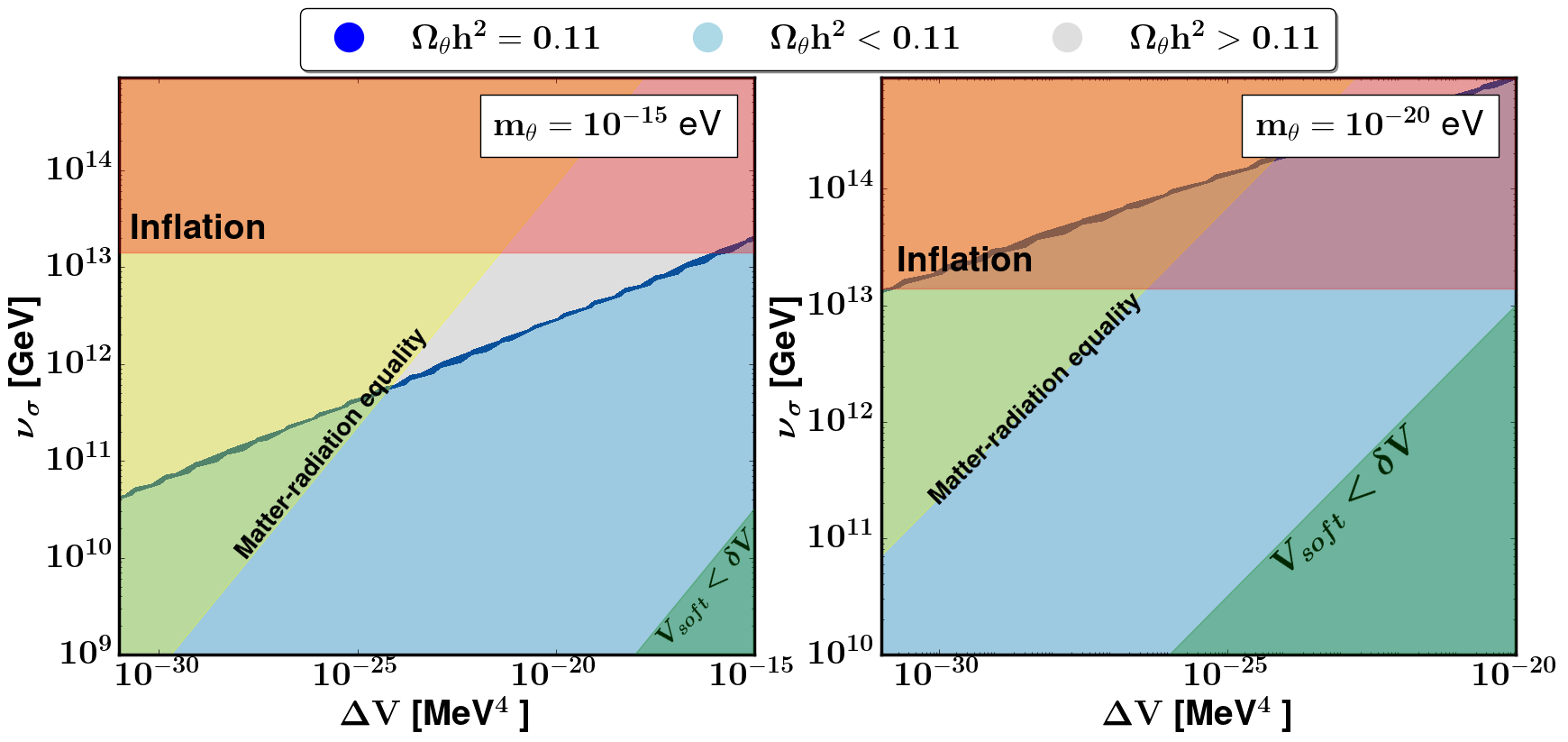}
    \caption{Viable parameter space in the plane ($\Delta V
    ,\nu_\sigma$) for the scenario where the SSB occurs after the end of inflation. The correct DM relic density is achieved along the blue line, the grey region represents $\Omega_\theta h^2 > 0.11$ while the light blue means $\Omega_\theta h^2 < 0.11$. The yellow region is excluded because the necessary requirement of DWs to decay before matter-radiation equality does not hold. The green region is excluded due to $V_\text{soft} < \delta V$}. The red region represents the end of inflation.
    \label{fig:abudanceII}
\end{figure}

In \autoref{fig:abudanceII} we present the allowed parameter space in the plane ($\Delta V,\nu_\sigma$) for the scenario where the SSB occurs after the end of inflation. The DM relic density that saturates the measured value, \textit{i.e.}~$\Omega_\theta h^2 = 0.11$, is achieved along the blue line. The grey region represents DM overdensity with $\Omega_\theta h^2 > 0.11$ while the light blue indicates $\Omega_\theta h^2 < 0.11$ where DM is underabundant. The yellow region is excluded as the necessary requirement for DW decay before the matter-radiation equality does not hold. The green region is excluded due to $V_\text{soft} < \delta V$. The red region represents the end of inflation. The left panel specializes to an ultralight scalar mass of $10^{-15}~\mathrm{eV}$, while the right panel considers $m_\theta=10^{-20}~\mathrm{eV}$. Notice that for the latter case there are no allowed points that could explain the measured relic abundance of DM. In turn, under the conditions of a post-inflationary scenario, oscillatons resulting from a clump of ultralight real scalars with mass close to $10^{-20}~\mathrm{eV}$ cannot be described in the context of a model where $\theta$ fully describes the DM component of the Universe, which contrasts with the pre-inflationary regime discussed for the scenario I.
\begin{figure}
    \centering
    \includegraphics[width=\columnwidth]{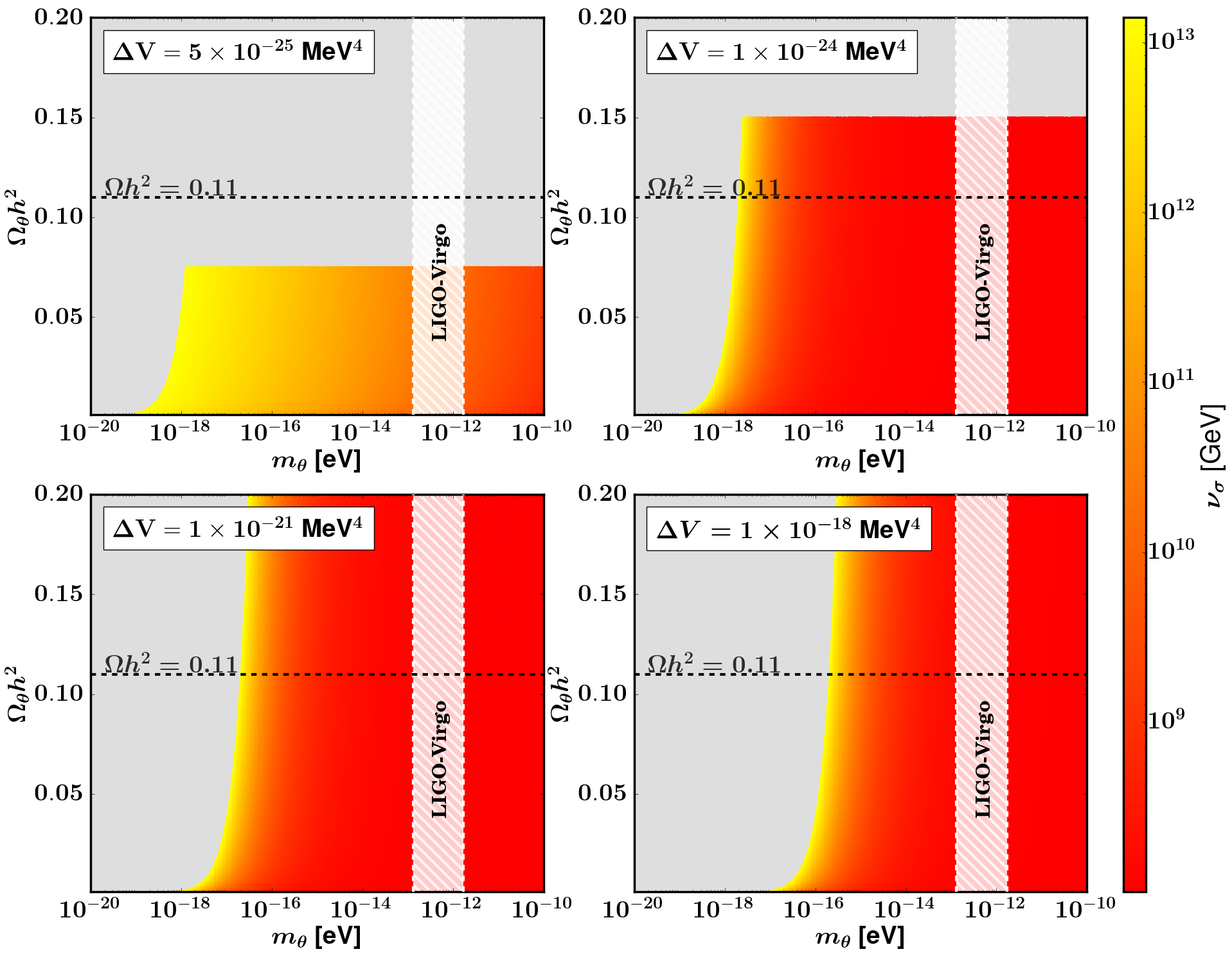}
    \caption{Viable parameter space in the plane ($m_\theta$, $\Omega_\theta h^2$), with the colour bar denoting $\nu_\sigma$, for the scenario where SSB occurs after the end of inflation. The white shaded region is excluded due to the constraints imposed by the LIGO-Virgo experiment in the searches for superradiance instabilities for a mass range $[1.3, 17] \times 10^{-13}$ eV \cite{Yuan:2022bem,Ng:2020ruv}.}
    \vspace{-0.4cm}
    \label{fig:abudanceIII}
\end{figure}

In \autoref{fig:abudanceIII} we present the allowed parameter space in the plane ($m_\theta$, $\Omega_\theta h^2$), with the $\U{G}$ breaking VEV, $\nu_\sigma$, represented in the color bar, for the scenario where SSB occurs after the end of inflation. The white shaded region indicates the excluded mass range due to searches for superradiance instabilities in the form of a SGWB at the LIGO-Virgo experiment \cite{Yuan:2022bem,Ng:2020ruv}. This figure corroborates the observation related to \autoref{fig:abudanceII} where it was verified than a mass of $10^{-20}~\mathrm{eV}$ is not compatible with a full description of DM. Indeed only masses larger than $\mathcal{O} (10^{-16})$ eV fulfil such a requirement as it is evident in \autoref{fig:abudanceIII}. Likewise, $\Delta V$ that has to be above $\mathcal{O} (10^{-24})~\mathrm{MeV}^4$. One must also note that $v_\sigma$ does not play a relevant role in the considered region.

\section{Gravitational waves from Domain Walls}\label{sec:pGW}

\begin{figure}
    \centering
    \includegraphics[width=\columnwidth]{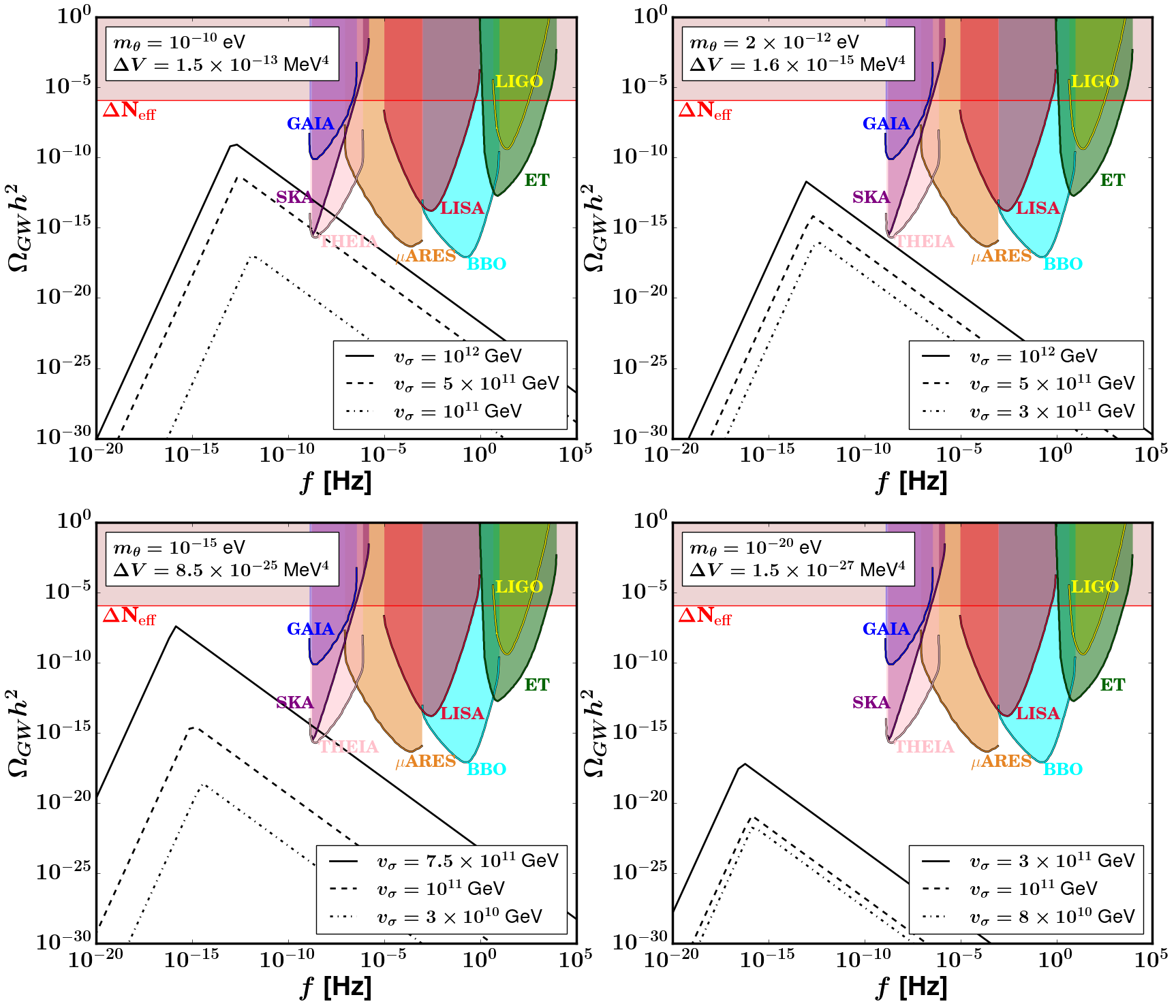}
    \caption{GW spectrum from DWs for different values of $\nu_\sigma$ (black curves) where each graph represents different values of $m_\theta$. The shaded region represents the experimental sensitivities of {\color{blue} GAIA}, {\color{Mulberry} SKA}, {\color{orange} $\mu$ARES}, {\color{BrickRed} LISA}, {\color{cyan} BBO},  {\color{Goldenrod}  LIGO}, {\color{pink} THEIA}  and {\color{ForestGreen} Einstein Telescope (ET)}. The red shaded region is excluded due to the $\Delta N_\text{eff}$ constraint.}
    \label{fig:pGW}
\end{figure}

The decay of DWs can generate not only DM but also a spectrum of SGWB \cite{Gelmini:2021yzu,Gelmini:2022nim,Hiramatsu:2012sc,Vaquero:2018tib}, expected with the characteristic peak frequency and amplitude features depending on the model's parameters. These can potentially be observed at current and future GW observatories, offering an unique probe of the Universe's earliest moments.

The SGWB amplitude at the peak frequency can be written as \cite{Kadota:2015dza, Hiramatsu:2013qaa,Saikawa:2017hiv}
\begin{equation}
\Omega_\text{GW} h^2 (t_0)_\text{peak} = 5.2 \times 10^{-20} A^4 \Tilde{\epsilon}_{gw} \left(\frac{g_*(T_\text{decay})}{10}\right)^{1/3}
\left(\frac{\sigma}{ \text{TeV}^3}\right)^4 \left(\frac{ \text{MeV}^4}{\Delta V}\right)^{2}\,,   
\end{equation}
where $\Tilde{\epsilon}_{gw} \simeq 0.7$ is the efficiency parameter and the peak frequency of the GW is given by
\begin{equation}
    f_\text{peak}(t_0) \simeq 3.99 \times 10^{-9} \text{ Hz } A^{-1/2} \left(\frac{ \text{TeV}^3}{\sigma}\right)^{1/2} \left(\frac{\Delta V}{ \text{MeV}^4}\right)^{1/2}
\end{equation}
The full spectrum is obtained by multiplying the peak amplitude by the spectral function as
\begin{equation}\label{eq:SpGW}
    \Omega_\text{GW} h^2 \simeq \Omega_\text{GW} h^2_\text{peak} \times
    \begin{cases}
      \left(\frac{f_\text{peak}}{f} \right), & \text{if}\ f > f_\text{peak} \\
      \left(\frac{f}{f_\text{peak}} \right)^3, & \text{if}\ f <
      f_\text{peak}\,.
    \end{cases}
\end{equation}

In \autoref{fig:pGW} we show examples of the SGWB given by \autoref{eq:SpGW} for different values of $\nu_\sigma$. The experimental sensitivities of GAIA \cite{Garcia-Bellido:2021zgu}, SKA \cite{Weltman:2018zrl}, LISA \cite{amaro2017laser}, $\mu$ARES \cite{Sesana:2019vho}, BBO \cite{Yagi:2011wg}, LIGO \cite{LIGOScientific:2014pky} and the Einstein Telescope (ET) \cite{Punturo:2010zz} are shown in the shaded regions. The energy density of a SGWB behaves like radiation \cite{Hiramatsu:2012sc}. Therefore, considering cosmological observations obtained from the PLANCK satellite and the limits on effective number of neutrino species ($\Delta N_\text{eff}$) derived from the CMB, one can extract an upper bound on the energy density amplitude of approximately $\Omega_\text{GW} h^2 \lesssim 10^{-6}$. The excluded region is represented by the red shaded horizontal band on the four panels. It is clear from our results that although the detection of a SGWB is not precluded, it is only possible for a restricted set of parameters and only at SKA and THEIA. Notice that the SGWB discussed here is of cosmological nature and not to be confused with that indicated in the vertical band of \autoref{fig:abudanceIII}, which is of an astrophysical origin.
\begin{figure}
    \centering
    \includegraphics[width=\columnwidth]{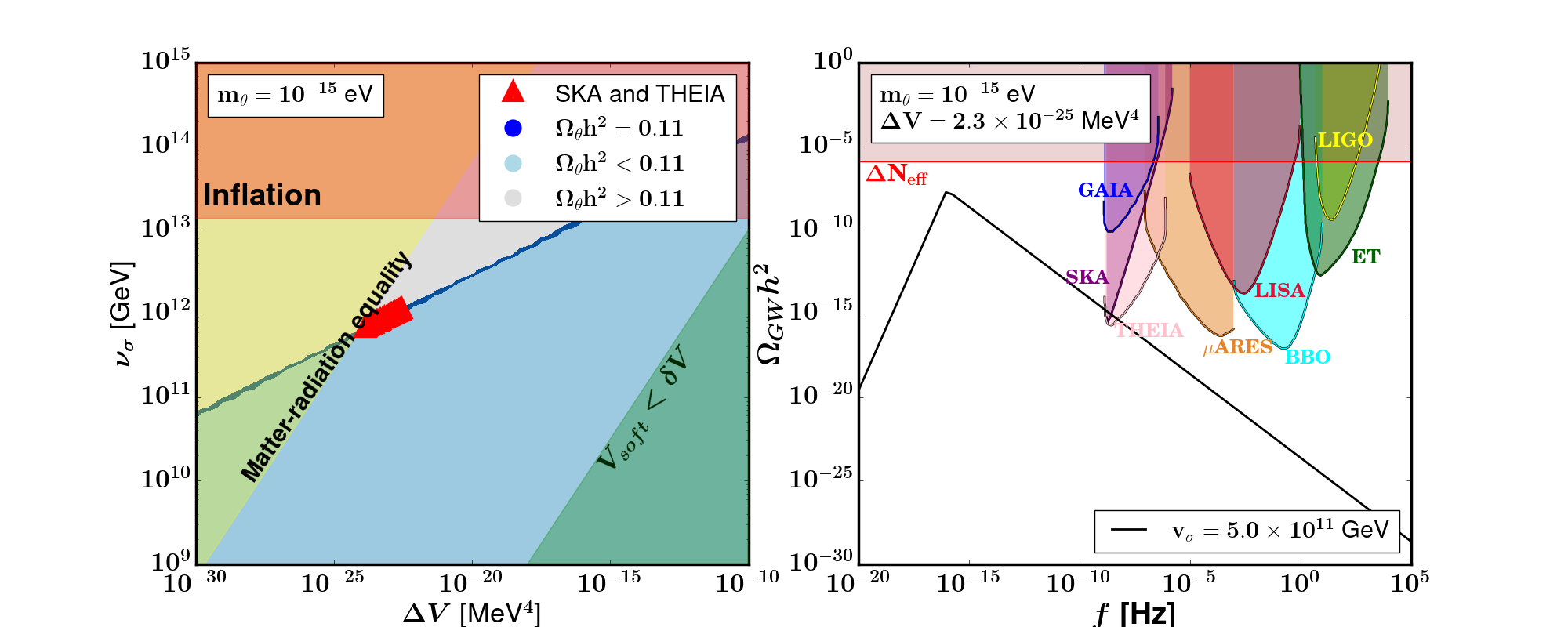}
    \caption{\textit{Left:} Viable parameter space in the plane $(\Delta V, \nu_\sigma)$ for $m_\theta = 10^{-15}$ eV as already presented in \autoref{fig:abudanceII}. The red triangles represent the set of parameters that provide the correct DM abundance and within reach of the experimental sensitivity of the SKA and THEIA experiments. \textit{Right:} Spectrum of GWs produced from DWs, where the black curves represents a choice of parameters corresponding to one red triangle in the left panel.}
    \label{fig:sensitivity}
\end{figure}

To conclude, in our analysis we revisit in the left panel of \autoref{fig:sensitivity} the viable parameter space in the plane $(\Delta V, \nu_\sigma)$ for $m_\theta = 10^{-15}$ eV, as already presented in \autoref{fig:abudanceII}, left panel. The added red triangles represent points that not only saturate the DM relic abundance but can also be tested by future GW experiments such as SKA and THEIA. In the right panel, we show the spectrum of GWs produced from DWs, where the black curves represent a choice of parameters that correspond to one of the red triangles in the left panel. The red triangles correspond to what we consider the most phenomenologically interesting solution of our analysis where the model describes DM in a post-inflationary regime, it can clump in the form of oscillatons and can be potentially tested in both astrophysical (LIGO-Virgo) and cosmological channels (SKA, THEIA).

\section{Conclusions}
\label{sec:conclusions}

We investigated a model that is a simple extension of the Standard Model by the addition of a complex singlet. When this new field acquires a non-zero VEV, a Nambu-Goldstone boson emerges. This NGB acquire mass through an additional soft breaking term in the potential becoming an pseudo Nambu Goldstone boson. We are focused on the scenario where the pNGB is an ultralight particle, in the mass range $\mathcal{O} \left(10^{-20} - 10^{-10} \right)$ eV, which makes it stable and therefore a good DM candidate.

Our motivation to explore this range of masses is primarily related to the strong gravity implications from the fact that ultralight relativistic bosons can clump and form objects with a compactness similar to that of black holes. With current experimental facilities, such objects are under active study and can be probed upon the measurement of an astrophysical SGWB emergent from superradiance instabilities.   

We have discussed the mechanisms of DM production relevant for the considered mass range. In particular, due to extremely suppressed couplings wit SM particles, DM is non-thermally produced through the misalignment mechanism and/or the decay of topological defects. However, when the breaking of the global $\U{G}$ symmetry occurs before the end of inflation there is no contribution from topological defects and it takes place predominately via misalignment. Indeed, we have noted that it is always possible to find a misalignment angle that saturates the observed DM relic abundance for the entire mass range relevant for strong gravity applications. For a post-inflationary regime, the main mechanism at play for DM generation is the decay of DWs. In this case, it is only possible to satisfy the entire DM relic density for ultralight scalar masses larger than approximately $10^{-16}~\mathrm{eV}$. Besides the known astrophysical SGWB observables at LIGO-Virgo frequencies, the model can also offer a cosmological SGWB potentially observable at SKA and THEIA.

With this article, one expects to provide valuable information to the strong gravity community, where for the case of oscillatons their solutions can be concretely/numerically mapped to viable DM scenarios in simple extensions of the SM of particle physics.

\section*{Acknowledgments}

The authors gratefully thank Carlos A.~R.~Herdeiro, Eugen Radu and Nicolas Sanchis-Gual for insightful discussions on strong gravity connections related to the presented work.
The authors also thank Fabrizio Rompineve for relevant comments related to domain walls.
This work was supported by the grants CERN/FIS-PAR/0021/2021, CERN/FIS-PAR/0019/2021, CERN/FIS-PAR/0024/2021, CERN/FIS-PAR/0025/2021 and PTDC/FIS-AST/3041/2020. A.P.M. is supported by the Center for Research and Development in Mathematics and Applications (CIDMA) through the Portuguese Foundation for Science and Technology (FCT - Fundação para a Ciência e a Tecnologia), references UIDB/04106/2020 and UIDP/04106/2020, and by national funds (OE), through FCT, I.P., in the scope of the framework contract foreseen in the numbers 4, 5 and 6 of the article 23, of the Decree-Law 57/2016, of August 29, changed by Law 57/2017, of July 19.
V.O. acknowledgment CAPES for financial support.
RS is supported by FCT under contracts UIDB/00618/2020, UIDP/00618/2020.
R.P.~is supported in part by the Swedish Research Council grant, contract number 2016-05996, as well as by the European Research Council (ERC) under the European Union's Horizon 2020 research and innovation programme (grant agreement No 668679).

\appendix

\bibliographystyle{JHEP}
\bibliography{biblio}
\end{document}